\documentclass[11pt,twoside]{article}
\usepackage{asp2010}
\usepackage{graphics}

\resetcounters

\begin{document}

\title{Ultra-compact binaries: relevance and role of Utrecht}
\author{Gijs Nelemans$^{1,2}$ and Lennart van Haaften$^1$
\affil{$^1$Department of Astrophysics/IMAPP, Radboud  University Nijmegen, P.O. Box 9010, NL-6500 GL, The Netherlands}
\affil{$^2$Institute for Astronomy, KU Leuven, Celestijnenlaan 200D, 3001 Leuven, 
Belgium}
}
\begin{abstract}
We present a short overview of the formation and evolution of
ultra-compact binaries. They are relevant to a surprisingly large
number of astrophysical phenomena (binary interactions, mass transfer
stability, explosive phenomena such as type Ia supernovae and
gravitational waves). 
\end{abstract}

\section{Introduction: what are ultra-compact binaries?}

Ultra-compact binaries refer to double stars with orbital periods
typically less than about one hour. They come in two general flavours:
detached systems and interacting binaries in which mass is transferred
from one star to the other. The limit of about one hour is derived
from a very simple argument, dating back to \citet[][see also
\citealt{ver97}]{pac71}. Each component in a binary has a radius at
most the size of its Roche lobe, which for a star of mass $M$ and
radius $R$ with a companion of mass $M_2$ can be approximated as
$R_{\rm L} = 0.46 a \frac{M^{1/3}}{(M + M_2)^{1/3}} \ge R$
\citep{pac71}. When combined with Kepler's third law to eliminate $a$
(and $M_2$!) using the orbital period $P$ this yields $M/R^3 \ge
\mathrm{const}/P^2$. Therefore, the smaller the period, the higher the
density of the objects and for periods smaller than one hour, the
densities are too high for main sequence stars. Ultra-compact binaries
therefore consist of evolved components: white dwarfs, neutron stars,
He stars or black holes, which makes them a class apart.

\section{Formation and evolution}

Because both components in ultra-compact binaries are evolved stars,
their formation is a probe of different uncertain processes in binary
evolution: common-envelope evolution, magnetic braking, massive star
evolution and NS/BH kicks. Three different formation
channels have been proposed \citep[see e.g.][]{nj10}:
\begin{description}
\item[In the WD channel] the primary evolves to become a
  compact object that then is engulfed in a common envelope with the
  secondary. After the common envelope the secondary becomes a WD and gravitational wave (GW) radiation brings the WD into contact
  with the primary at a period of a few minutes. If the mass transfer
  is stable (see \S \ref{stab}), the orbit will widen again (Fig.~1).
\item[In the He star channel] the evolution is largely the same,
  except that the core of the secondary after the common envelope is a
  core He burning star that starts mass transfer to the primary at
  periods of several tens of minutes and first evolves to shorter
  periods, until a minimum of around 10 minutes is reached and then
  turns around.
\item[In the evolved main-sequence channel] the secondary fills its
  Roche lobe to the compact object when it just reaches the end of the
  main sequence. It first evolves as a normal low-mass X-ray binary or
  cataclysmic variable, until the H depleted core of the star becomes
  exposed. The system essentially merges onto the He star channel,
  although in many cases the transferred material still shows traces
  of H \citep[e.g.][]{prp02}. Work done in Utrecht showed that in
  order to get to very short orbital periods, fine tuning of the
  initial binary parameters as well as strong magnetic braking is
  needed \citep{svp05a}.
\end{description}

\section{Stability at the onset mass transfer}
\label{stab}

\begin{figure}
\includegraphics[width=0.5\textwidth,clip=]{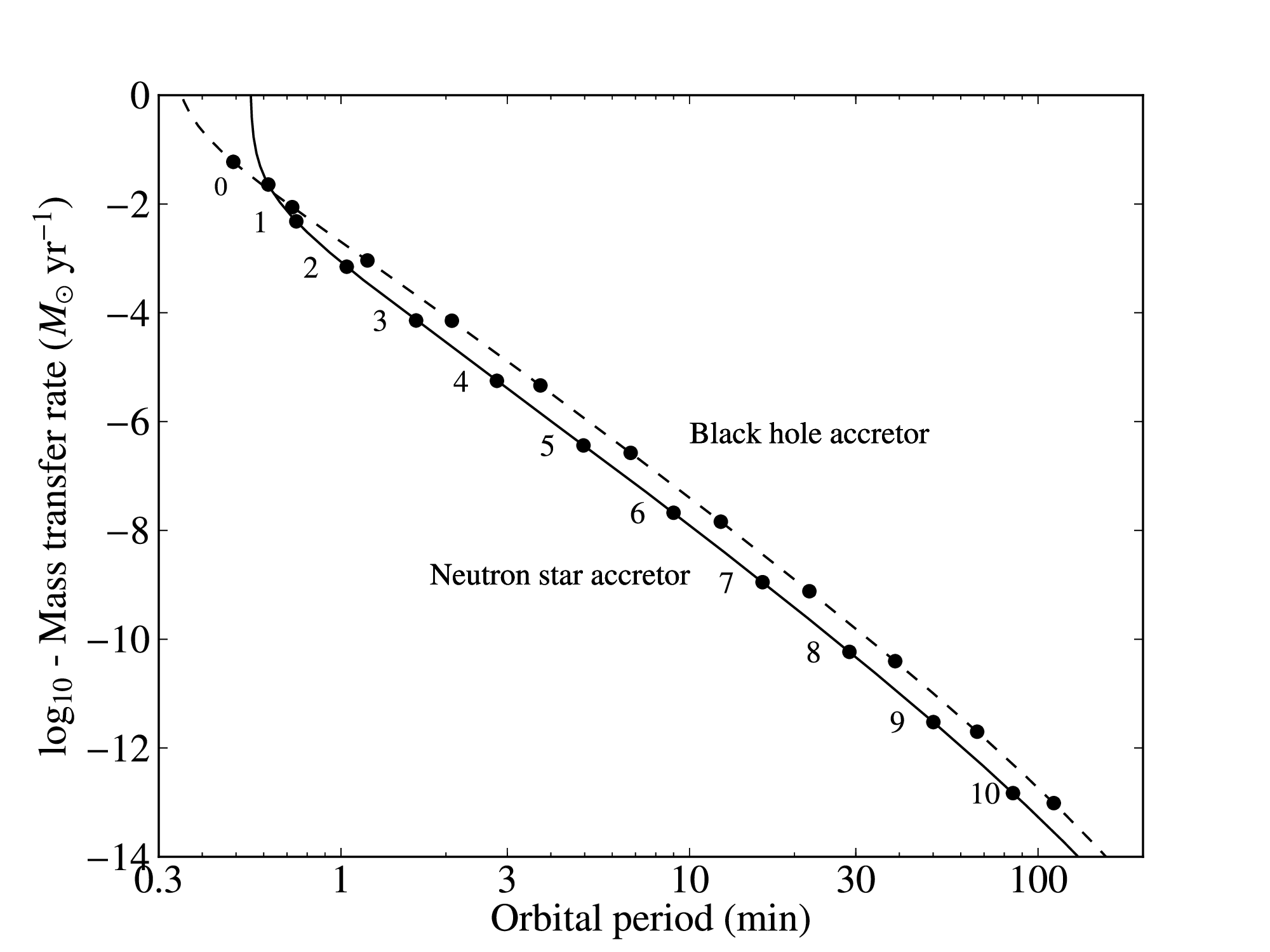}
\includegraphics[width=0.5\textwidth,clip=]{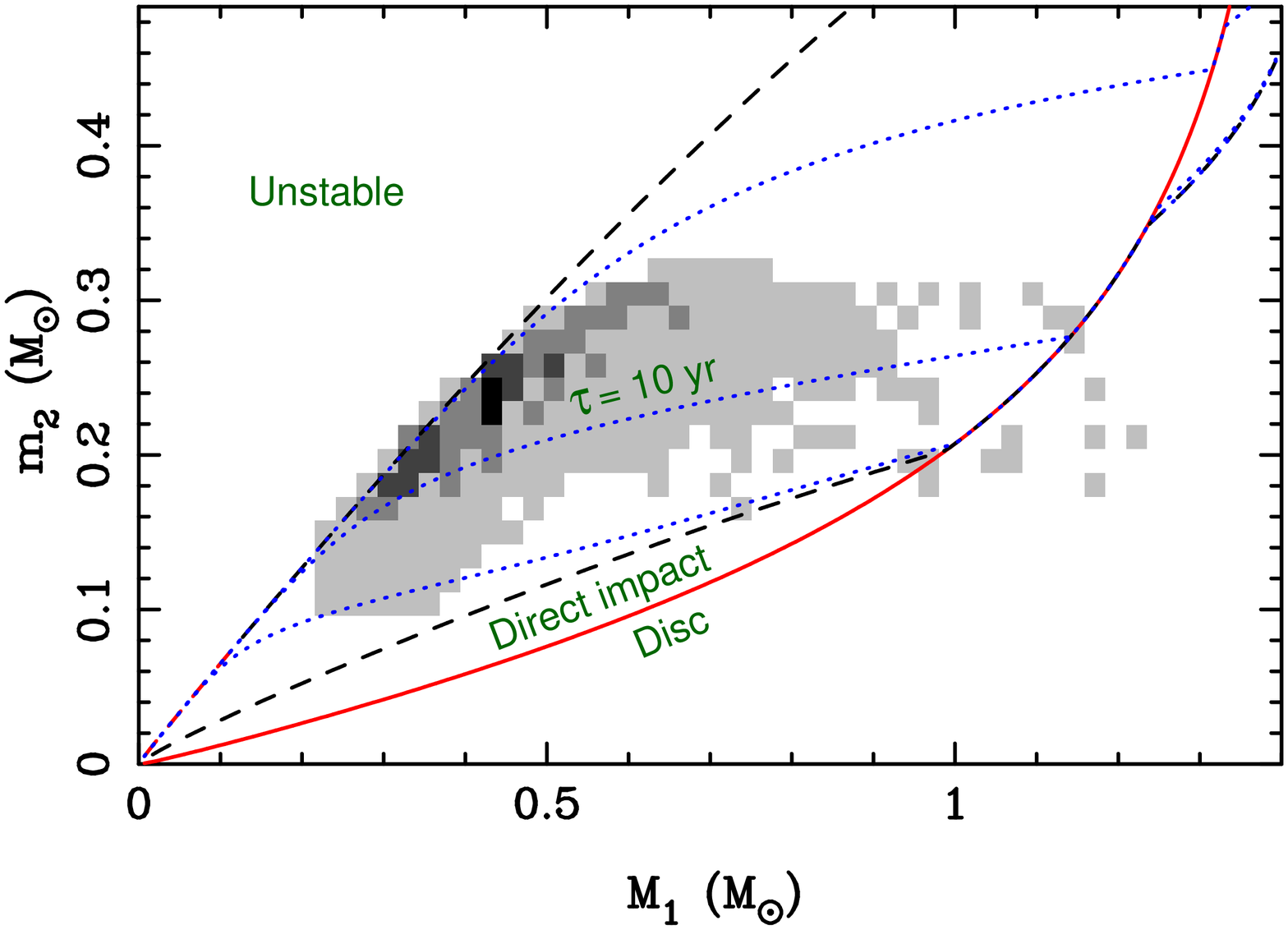}
\label{fig:PMdot}
\caption[]{Left: Mass transfer evolution as a function of orbital
  period for UCXBs, the dots indicating the logarithm of the age of
  the system in years. From \citet{hnvwk12}. Right: stability
  regions for the onset of mass transfer between two WD. For
  direct impact accretion (middle region with most expected systems)
  stability depends on tidal coupling. From \citet{mns04}.}
\end{figure}

The onset of mass transfer is an extremely interesting phase in the
case of a WD donor. Because WD have an inverted mass-radius relation,
the donor will always be the lower mass object. The stability
criterion ($R = R_\mathrm{L}$ and $\dot{R} = \dot{R_\mathrm{L}}$ at
all times) immediately gives an upper limit to the mass ratio of the
system \citep[see][]{hnvwk12}.  But even if in principle the mass
transfer can be stable, the time scales of the evolution due to GW
emission are short and the mass transfer rates high (Fig.~1). For NS
(and BH) accretors this means the mass transfer is highly
super-Eddington. Assuming radiation liberated by accretion is used to
expel the excess matter from the inner Lagrangian point (L1) one can
derive a (likely) upper limit on the mass transfer that can still be
stable \citep{kr99}. Because the mass transfer scales with the donor
mass this puts an upper limit to the initial mass of a WD donor of
$0.38\ M_{\odot}$. Therefore He core WD can survive the onset of mass
transfer more easily.

For WD accretors another interesting issue arises: at
these very short periods, the accretor is relatively big compared to
the orbit and in many cases the mass stream from L1 will hit the
surface of the accretor directly (so-called direct impact, see
Fig.~1). This complicates the stability analysis of mass transfer, as
the usual condition that the angular momentum that the donor loses to
the stream is stored in the disk that forms and then via tides put
back in the orbit \citep[and thus can effectively be ignored,
see][]{vr88} is not obvious anymore. This could seriously destabilise
the mass transfer (see Fig.~1), unless there is strong tidal
interaction between the orbit and the spinning accretor (which is
unknown, \citealp{mns04}).

\section{Explosive phenomena}

The evolution around the onset of mass transfer can give rise to
various explosive phenomena: if the mass transfer is unstable the two
stars will merge. In the case of two WDs with enough mass, this has
been proposed as a possible origin of type Ia supernovae. For lower
mass WD the outcomes are more likely R CrB or sdB stars
\citep{web84}. Less work has been done on the merger of a WD with a
NS, but recently \citet{metzger12} suggested they may produce
sub-luminous supernovae.

Even if the mass transfer is stable, there may still be fire
works. The two known classes of objects are ultra-compact X-ray
binaries (UCXBs; a NS accreting He or a C/O mixture, see
\citet{hnvwk12} and AM CVn stars (a WD accreting He,
\citealp[see][]{solheim10}).  The mass transfer is driven by GW
emission and is characterised by rapidly dropping mass transfer rates
(see Fig.~1). In AM CVn stars the mass transfer rate of He onto a WD
in the early phase is high enough to produce He novae. At periods of
around 10 min, the mass transfer rate has dropped enough that the
ignition mass becomes so large that the layer is degenerate enough for
the burning to proceed on a dynamical timescale, leading to a
thermonuclear explosion, observable as a sub-luminous supernova,
called .Ia supernova \citep{bswn07}.  With the advent of wide-field
variability surveys such as Pan-Starrs and PFT, there are good chances
of finding these supernovae. Indeed several new types have been found
\citep[e.g.][]{kb12}, but it is (yet) unclear which observed types
belong to which theories!

\vspace*{-0.3cm}
\section{Late time evolution}

\begin{figure}
\includegraphics[width=0.5\textwidth]{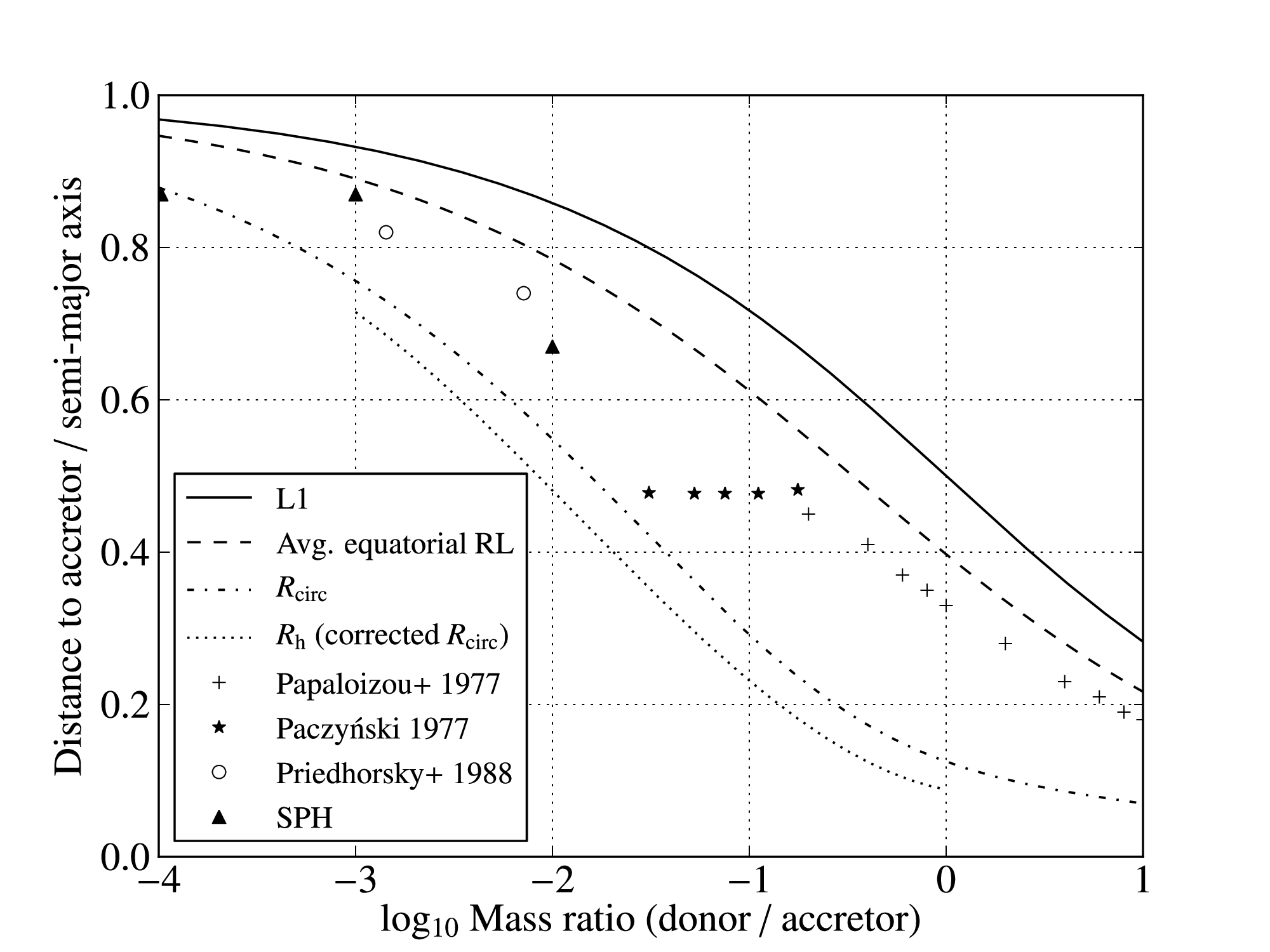}
\includegraphics[width=0.5\textwidth]{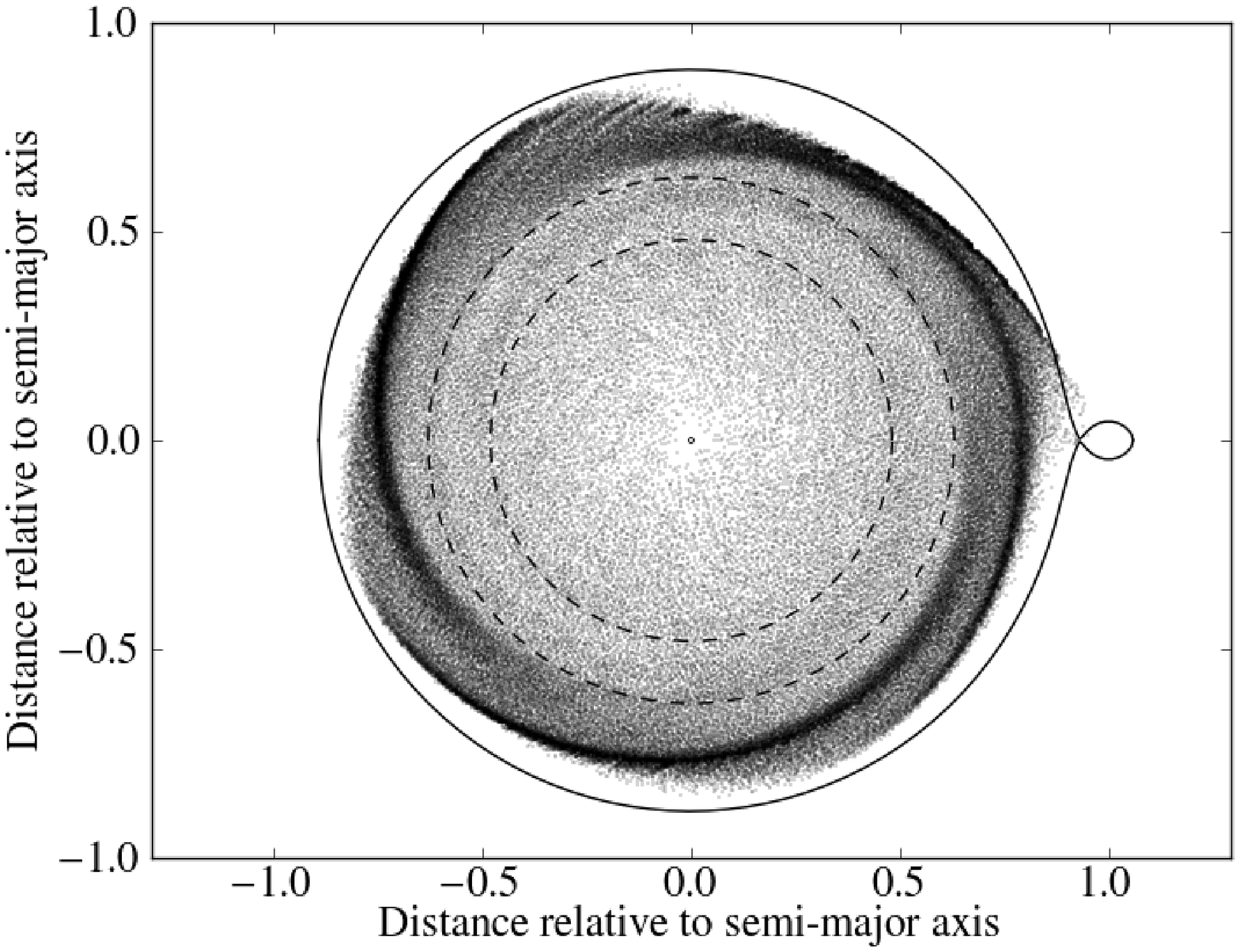}
\label{fig:late}
\caption[]{Left: radii of Roche lobe, circularisation radius ($R_\mathrm{h}$)
  and outer edge of the disk according to several authors as function
  of mass ratio. Right: SPH calculation of a disk in an extreme mass
  ratio case. From \citet{hnvwk12}.}
\end{figure}

If the systems survive the early violent evolution, the mass transfer
rate continues to go down (Fig.~1). The mass ratio becomes very
extreme ($< 0.01$), and the disk has to grow larger and larger to
redistribute the angular momentum and it has been suggested that this
would be impossible, leading to an instability. However, \citet{pv88}
already showed that that was likely not the case and recent SPH
calculations indeed confirm that the disk becomes large and the system
survives (see Fig.~2 and \citet{hnvwk12}), leading to a large pile up
of systems at long orbital periods. For AM CVn systems indeed many
long period systems are found \citep[see][]{solheim10}.  However, this
is not the case for UCXBs. We explored the options to explain this and
conclude that either the systems are invisible most of the time
(e.g. due to thermal instabilities in the disk), or the evolution is
sped up by X-ray induced wind mass loss from the donor, which is also
consistent with the average X-ray luminosity of the few long-period
UCXBs and the idea that the millisecond radio pulsar system PSR
J1719$-$1438 is a remnant of UCXB evolution (see
\citealt{bailes11,hnvj12}).

\section{Gravitational waves}

The evolution of ultra-compact binaries is largely driven by
GW radiation. It is therefore maybe not
unexpected that they are important GW sources. Indeed, space detectors
such as the eLISA mission \citep{amaro12} should be able
to detect several of the known AM CVn stars and a detached double WD
with a period of only 12 min \citep{brown11}. They thus are
guaranteed or verification sources. More interesting, eLISA should
detect several thousand new ultra-compact binaries most with periods
shorter than 10 minutes! It should detect \emph{all} NS star (and
BH) binaries with periods shorter than 35 min in the whole
Galaxy.

\acknowledgements As SIU alumni we would like to thank all people at
SIU for our education and training and for fruitful collaborations
over the years.


\end{document}